\documentclass[%
 reprint, amsmath,amssymb, aps]{revtex4-2}
\usepackage{graphicx}% Include figure files
\usepackage{dcolumn}% Align table columns on decimal point
\usepackage{bm}% bold math
\usepackage{physics}
\usepackage{amssymb, amsmath, amsmath, amsfonts}

\begin{document}

\preprint{ARXIV/2112.15503}

\title{Space as relation}% Force line breaks with \\

\author{Marcello Poletti}
 \email{epomops@gmail.com}
\affiliation{%
 San Giovanni Bianco, Italy
}%
\date{\today}% It is always \today, today,

\begin{abstract}
Here we will discuss the philosophical differences between an approach to the deep nature of physical space based on the concept of coordinates and one based on the concept of relation.  The philosophical superiority of the second approach will be analysed and maintained, attempting to bring it to its extreme consequences.
We will propose considering the concept of spatial dimension completely superfluous in favour of an idea of space as a pure relational structure, capable of including within itself the information necessary to define the properties of the things in the world.

\end{abstract}

\maketitle

%\tableofcontents

\section{Introduction}
In his "Popular Exposition"\cite{B1}, published just a year after the fundamental articles of 1915, Einstein describes an attempt by Ernst Mach to solve the problem of inertia by rethinking space in a purely relational form, based on the concept of distance rather than position.

References to this possibility are found in many authors, from Poincaré\cite{B2} to Rovelli and Vidotto\cite{b3}.

Classical geometry typically studies the relationships between geometric entities such as points, segments, and lines. The concept of position (or coordinate) has been imposed since geometry became analytic with the famous works of Descartes.

Analytic geometry is capable of transforming properties of the pair of entities $\{A,B\}$ into atomic pseudo-properties of $A$ and $B$, forcing a kind of reductionism through the introduction of arbitrary properties and functions through which relational properties can be reconstructed.

In particular, the relational property "A and B are d apart" is transformed into the reduced properties:

\begin{itemize}
	\item $A$ has coordinates $A_i$
	\item $B$ has coordinates $B_i$
\end{itemize}
From these properties is possible to reconstruct the distance $d$ through an appropriate function:
\begin{equation}\label{dist}
	d(A,B)=\sqrt{\sum{(A_i-B_i)^2}}
\end{equation}
The properties $A_i$ and $B_i$ turn out to be completely arbitrary, and can be freely chosen within a series of transformations (rototranslations) that keep  \ref{dist} invariant; for all intents and purposes, these properties act as phases whose proper value is arbitrary and whose only concrete effect is to enter into a relation (typically a difference) with other phases of the same type.

The function selected for the distance is such as to automatically guarantee that triangular inequalities are respected for every possible choice of coordinates, and this is extraordinarily effective in freeing the geometry from annoying constructability constraints.

This approach, which we will call \textit{positional} here, introduces an additional fundamental parameter: the number of coordinates to be used for each point, the \textit{number of dimensions}.

This value is equally arbitrary in the reconstruction of the distance $(A,B)$, while it assumes a central role in the case of considering more than just two points. In particular, this value plays the role of a constraint with respect to the possible distance values allowed; in an m-dimensional space for $m+1$ points only, it is possible to define every possible combination of distance, respecting the triangular inequalities; additional points will be (automatically) constrained so that some distances, even if compatible with the triangular inequalities, are not allowed.

Alternatively, it is possible to define the geometry of $n$ points through their reciprocal $n(n-1)/2$ distance relations, subject to the complex constraints given by the triangular inequalities. This approach, which we will call \textit{relational}, presents no symmetries and has no dimension parameters. The relational approach is \textbf{not} equivalent to the positional approach, because the constraints imposed by dimensionality do not apply to it. In a relational geometry, by definition, \textbf{every possible distance is allowed, respecting only the triangular inequalities}. In this sense, the relational approach is broader than the positional one.

\section{Non-Euclidean Geometry}
In non-Euclidean geometry, the positional approach undergoes important updates.

In addition to $A_i$ coordinates, the concept of metric tensor $g_{ab}$ is introduced, defining the shape of the space through the fundamental relation which extends the \ref{dist}
\begin{equation*}
	ds^2=g_{ab}dx^adx^b
\end{equation*}
The action of $g_{ab}$ is to make relationships of distance more elastic by reducing the role of the constraints given by the dimensions; a curved variety allows distances typically not definable in a Euclidean space.

It is natural to wonder to exactly what extent these constraints are reduced. The answer, curiously rarely emphasized in geometry textbooks, is that such constraints disappear altogether in spaces of dimension 3 or higher [Appendix \ref{Appendix}].

The non-Euclidean positional approach, in relation to reciprocal distances, is equivalent to the relational approach.

Viewed retrospectively, then, non-Euclidean geometry is none other than a way to free the positional approach from its dimensional constraints. The positional approach is finally equivalent to the relational approach for varieties of dimension at least 3. Conversely, the relational approach spontaneously conduces to non-Euclidean spaces of dimension at least 3; Mach's intuition was therefore correct: the relational approach has in itself properties equivalent to the concept of curvature and it was precisely these properties that led Einstein to the solution of the problem of inertia.

At this point, however, one wonders if the advantages introduced by the positional approach, particularly the automatic validity of triangular inequalities, continues to outweigh its costs, in particular:

\begin{itemize}
	\item The complexity presented by the $g_{ab}$ tensor
	\item The presence of gauges due to the arbitrariness of the coordinates/phases.
	\item The presence of the ambiguous parameter related to the number of dimensions.
\end{itemize}
\section{Metric spaces}
The formalization of the positional approach leads to the concept of normed vector space; the formalization of the relational approach leads, in the first instance, to the weaker concept of metric space.

On the one hand, vector spaces have shown themselves to be extraordinarily effective in a great many areas even beyond the modelling of physical space; on the other hand, metric spaces turn out to be too poor as possible substitutes. In particular, the former allow us to easily define all essential geometric sizes, angles, areas, volumes, etc., while the latter do not.

A metric space is essentially a set over which distances are defined, respecting triangular inequalities. This definition is ultimately too weak, as it does not include the concept of path or beetwinnes\cite{B4}.

Intuitively, what characterizes the distance relation in physical space between points $A$ and $B$ is precisely the existence of paths between $A$ and $B$, “distance” emerging a posteriori as a search for the minimum of the path lengths. In simple terms, $A$ and $B$ are 6 steps apart in the sense that it is, at a minimum, possible to travel 6 steps to get from $A$ to $B$, and this implies that point C also exists at step 1, point $D$ at step 2, and so on.

Since metric spaces do not contain the concept of path, they do not contain the concept of topology, nor the concepts of angle, area, volume, etc.

There are many alternatives oriented toward solving this difficulty: Alexandrov spaces\cite{B5}, length spaces\cite{B6}, topological spaces\cite{B7}, angle spaces\cite{B8}, proximity spaces\cite{B9}, etc.

These last are particularly attractive with respect to instances of natural philosophy, basing the spatial relationship on the classical concept of locality: $A$ is proximate to $B$, or adjacent to $B$ or, more simply near $B$, while $C$ is distant and reachable through a chain of proximity.

\section{Proximity spaces}
A proximity space $(X,\rho)$ is defined as a set $X$ equipped with a function with domain in $X$ and co-domain on the set of parts of $X$, $\rho(X)\to\mathcal{P}(X)$, such that
\begin{equation*}
	\begin{cases}
		A\notin\rho(A) \\
		A\in\rho(B)\iff B\in\rho(A)
	\end{cases}
\end{equation*}	
This definition of proximity space is remarkably simple, inherently discrete, and entirely sufficient for the purposes of the reflections made here.

A proximity chain is defined as a succession of points $\{P_i\}$ such that $P_i\in\rho(P_{i+1})$, empty successions or successions containing only one point are conventionally included.

A path between $A$ and $B$ is defined as a finite proximity chain that starts at $A$ and ends at $B$.
\begin{equation*}
	P^A_B = \begin{cases}
		\{P_i\}_{i=1}^n \\
		P_1=A \\
		P_n=B
	\end{cases}
\end{equation*}			
The length of the path between $A$ and $B$ is defined as the quantity:
\begin{equation*}
	|P^A_B| = Card(\{P_i\}_{i=1}^n)-1=n-1
\end{equation*}
The proximity space $(X,\rho)$ will be considered complete if, for each pair of points $A$, $B$, there exists a path between $A$ and $B$; it will be considered finite if $Card(X)<\aleph_0$ and it will be considered locally finite if $\forall A\in X,Card(\rho(A))<\aleph_0$

On a proximity space it turns out that the metric is clearly defined:
\begin{equation*}
	d(A,B)=min(|P_A^B |)
\end{equation*}
Note how in a proximity space the concept of distance is in a certain sense accidental: distance is a minimum of a larger condition, the length of the paths, and this complex definition leads, again accidentally, to the condition of triangular inequalities.

A finite proximity space substantially coincides with a graph on which the metric induced by the minimal paths is defined; a locally finite proximity space on the other hand has the appearance of a possibly infinitely extended graph.
\section{Dimensions}
There are several cutting-edge theories that take an approach to the nature of the space in various aspects in line with what is discussed here; we think in particular of LQG\cite{B15}, Causal Sets\cite{B14}, and the Wolfram Physic Project\cite{B16} (WPP). 

In all these cases, however, we observe a forcing to make the chosen structure "adapt" to a parameter that is itself absent in the base setting, the number of dimensions.

In the WPP manifesto\cite{B10} we read, for example:

\begin{quote}\textit{
I'm frankly amazed at how much we've been able to figure out just from the general structure of our models. But to get a final fundamental theory of physics we've still got to find a specific rule. A rule that gives us 3 (or so) dimensions of space, …
}
\end{quote}
And Fay Dowker, writes, in relation to Causal Sets\cite{B11}:
\begin{quote}\textit{
If spacetime is a causet, however, coarse graining it at different scales (i.e. with different
deletion probabilities p) gives rise to different causets which may have continuum
approximations with different topologies, including different dimensions.}
\end{quote}
The impression is that these authors are searching for a relational space which, at least at a certain scale, adequately emulates a three-dimensional positional approach, but as we can see, this is essentially useless. The number of dimensions can be considered entirely superabundant, emergent if you will, an unwelcome remnant of the positional approach, a necessary cost of vector formalism and its extraordinary effectiveness.

The relational approach under discussion does not indicate such a parameter, yet it is quite evident that this parameter has a role in physical space: it is possible to arrange three pencils so that they are mutually perpendicular, and it is not possible to do it with four.

One possible solution to this paradox is that the concept of "pencil" be closely related to the metric properties themselves, and this is exactly what happens in general relativity. 

Wheeler wrote\cite{B12}:
\begin{quote}\textit{
Could the metric continuum be a magical medium, which folded here in a given way represents a gravitational field, waved there in another way describes an electromagnetic field, and twisted locally describes a persistent concentration of mass energy?
}\end{quote}
This question is partially answered by Einstein's field equations:
\begin{equation*}
	R_{\mu\nu}-\frac{1}{2}g_{\mu\nu}+\Lambda Rg_{\mu\nu}=\frac{8\pi G}{c^4}T_{\mu\nu}
\end{equation*}
which, usually read as "matter bends space-time", can just as well be read as "matter IS a fold of space-time".

If "pencil" is by definition the curvature of the 3D categorization of an inherently dimensionless space, then the paradox dissolves, "pencil" is a 3D object by definition.

This approach also allows us to reread some overdimensional theories from a new perspective. In particular, the Kaluza hypothesis can be considered completely legitimate in this perspective, redefining the electromagnetic field as a categorization of the same space in 4D, responding positively to Wheeler’s provocation for this field as well, and without any need to assume "compactifications" of the exceeding dimension, such spaces being purely emergent, categorizations of an underlying dimensionless space.
\section{Conclusions}
The historical and philosophical analysis of the concept of space conducted here compares the concept of Cartesian space - container of geometric entities - with the temptation that emerged many times in the history of scientific thought: giving up a containing background in favour of a geometry based purely on reciprocal relations.

From this emerges an idea of space that leads in the direction of Wheeler's fascinating proposal, a space that itself defines its content as a function of its own metric, which defines the real as the emergence of metric properties.

In a recent video\cite{B13}, David Gross said  \begin{quote}\textit{"...there are two facts that we know about nature: one is there’s matter, like us..."}\end{quote} The approach discussed here denies this statement, relegating matter to a pure 3D categorization of these properties.

Space - even in its traditional Euclidean view - is a purely relational structure, in which a marvellous recursion makes it so that the related entities within it, the points, have no property except to be able to enter into relation with other points. This structure is capable of holding local information by itself, encoded in the metric that is created (GR).

Note the philosophical difference between a vector of bits, objects capable of “containing” a value $\{0,1\}$ and implicitly ordered (otherwise the information contained in it is null), and the structure under discussion (for simplicity think of a graph) in which there is “nothing added” to the state of proximity of the points.

In this fascinating scenario, it seems necessary to also give up that parameter of dimension, so central to the reflections within physics of the last 50 years, to relegate it to a pure emergent value.

From the ontological perspective, it is frankly difficult to thoroughly estimate how much what is discussed has to do with “what is” and how much has instead to do with “what describes what is”, but even in this weaker second version, I believe that the relational approach can be a mental beacon of reference.  This idea essentially endorses the LQG, Causal Set, WPP-type approaches to the new physics (which not coincidentally exploit graph theory in one way or another), inviting promoters to avoid forcing - such as the search for particularly ad hoc spaces or the introduction of additional parameters (the addition of matter, the addition of properties of points, the addition of properties on the proximity, etc.) - by treating the spatial structure itself as a container for the necessary information by acting purely on the metric properties, following Wheeler's proposal and provocation to the end.

\begin{acknowledgments}
To my brother Giacomo, prince of calembour.
\end{acknowledgments}

\appendix

\section{}\label{Appendix}
Let us consider $n$ points and their $n(n-1)/2$ freely chosen distances, respecting only the triangular inequalities. Let $d$ be the maximum of such distance.

Now arbitrarily choose $n$ points in a Euclidean space inside a sphere of diameter less than $d$ and connect each pair with a path of length equal to the distance, so that two paths never cross. This is always possible in at least 3D space. 

Then deform the metric at each point that does not belong to a path, so that these paths become geodesics. The manifold finally obtained contains the metric of the $n$ starting points.
\nocite{*}

\bibliography{Eng}% Produces the bibliography via BibTeX.

\end{document}